# Angle-dependent magnetoresistance as a sensitive probe of the charge density wave in quasi-one-dimensional semimetal Ta$_2$NiSe$_7$

Jiaming He,[1] Libin Wen,[1] Yueshen Wu,[1] Jinyu Liu,[2] Guoxiong Tang,[1] Yusen Yang,[1] Hui Xing,[1,†] Zhiqiang Mao,[2] Hong Sun,[1] Ying Liu[1,3,§]

[1]Key Laboratory of Artificial Structures and Quantum Control and Shanghai Center for Complex Physics, School of Physics and Astronomy, Shanghai Jiao Tong University, Shanghai 200240, China
[2]Department of Physics and Engineering Physics, Tulane University, New Orleans, Louisiana 70118, USA
[3]Department of Physics and Materials Research Institute, Pennsylvania State University, University Park, Pennsylvania 16802, USA

†Email: huixing@sjtu.edu.cn
§Email: yxl15@psu.edu

**Abstract:**

The behavior of charge density wave (CDW) in an external magnetic field is dictated by both orbital and Pauli (Zeeman) effects. A quasi-one-dimensional (Q1D) system features Q1D Fermi surfaces that allow these effects to be distinguished, which in turn can provide sensitive probe to the underlying electronic states. Here we studied the field dependence of an incommensurate CDW in a transition-metal chalcogenide Ta$_2$NiSe$_7$ with a Q1D chain structure. The angle-dependent magnetoresistance (MR) is found to be very sensitive to the relative orientation between the magnetic field and the chain direction. With an applied current fixed along the *b* axis (the chain direction), the angle-dependent MR shows a striking change of the symmetry below $T_{CDW}$ only for a rotating magnetic field in the *ac* plane. In contrast, the symmetry axis remains unchanged for other configurations (*H* in *ab* and *bc* plane). The orbital effect conforms to the lattice symmetry, while Pauli effect in the form of $\mu_B B / \hbar v_F$ can be responsible for such symmetry change, provided that the Fermi velocity $v_F$ is significantly anisotropic and the nesting vector changes in a magnetic field, which is corroborated by our first-principles calculations. Our results show that the angle-dependent MR is a sensitive transport probe of CDW and can be useful for the study of low-dimensional systems in general.

Condensed matter systems with low dimensionality have demonstrated great potential by hosting rich and exotic physics[1,2] which allows an alternative and fascinating route for exploring exotic phenomena, including the recent discovery of valley dependent transport[3] and superconductivity[4] in $MoS_2$, extremely large magnetoresistance (MR) in $WTe_2$[5], and the topological phases.[6] These systems often display electronic superstructures with charge/spin density waves (CDW/SDW), which form through the Peierls transition due to the instability of the Q1D and 2D structure against the reduction of electronic energy.[7] As a representative electronic superstructure, CDW concerns a state with a two-fermion condensate involving a coherent superposition of electron and hole pairs. The intimate relation between CDW/SDW and superconductivity,[8-10] and possible connection to quantum phases such as Luttinger liquid[11,12] are among the major research efforts. These facts raise special interest in the study of CDW in Q1D metallic systems for exploring novel behaviors, which often involve external stimuli, including chemical doping/intercalation,[13-15] electric field[16,17] and external pressure.[18,19] In comparison, the effect of external magnetic field is of fundamental interests. The magnetic field effect on CDW involves the orbital effect, Pauli effect and several other material-specific properties such as the inter-chain hopping, and Fermi surface nesting vectors, etc. For Q1D CDW, these complications make the CDW phase diagram in a magnetic field rather complex.[20] With its intimate connection to quantum phases with fingerprints on magentotransport[12] and the quantized phase of field-induced CDWs/SDWs,[20-22] the study on the magnetic field dependence of various CDW systems is of particular interest.

The focus of this work is a Q1D ternary transition metal chalcogenide $Ta_2NiSe_7$ with an incommensurate CDW[23]. The transport properties of $Ta_2NiSe_7$ and the relation to the underlying electronic states were studied in our earlier work.[24] In addition to a small kink in the temperature dependence of resistivity corresponding to the CDW,[23] a clear change of curvature in the field dependence of MR is observed upon entering the CDW state, as a result of the CDW gap opening mainly on the hole-like $p$ state from Se atoms. Nevertheless, it is rather surprising to note that so far no signature in electron transport has reflected the Q1D character of the system. Here we find that the angle-dependent MR in $Ta_2NiSe_7$ shows a striking change of the symmetry axis below the CDW transition only for a rotating magnetic field in the $ac$ plane and current along the $b$ axis, while for other configurations ($H$ in $bc$ plane or $H \perp c$) the symmetry axis remains unchanged. We propose that the Pauli effect is responsible for such symmetry change given that the Fermi velocity $v_F$ is significantly anisotropic on the relevant surface, and the Fermi surface nesting vector changes in a magnetic field. Our first-principles calculations revealed consistently dominant flat Fermi surfaces with anisotropic Fermi velocities, and finite inter-chain coupling which can facilitate the shift of CDW nesting wave vector in a magnetic field. The angle-dependent MR provides a sensitive and convenient transport probe of CDW and revealed the Q1D nature of the system, and may find more application in the study of low-dimensional systems.

High-quality Ta$_2$NiSe$_7$ single crystals were prepared using the flux method. X-ray diffraction (XRD) was performed on a Bruker D8 diffractometer. XRD showed a space group of (C2/m) and lattice constants of $a$ = 13.84 Å, $b$ = 3.48 Å, $c$ = 18.60 Å, $α$ = $γ$ = 90°, $β$ = 108.8°, consistent with a previous report.[25] Single crystal morphology and elemental analyses were carried out by scanning electron microscopy and energy dispersive X-ray spectroscopy, respectively. Crystals used in our measurements are from the same batch used in our previous study.[24] They are highly selected, with a residual resistance ratio greater than 7, which is the highest among those reported in the literature. The typical size of the crystal is 1000 × 20 × 10 μm$^3$. All resistance was measured with current of 200 μA applied along the $b$ axis (chain direction), by using the standard four terminal method in a Quantum Design Physical Property Measurement System with a 14 T magnet, with a rotator for controlling the relative orientation between the magnetic field and crystal. No current amplitude effect was observed. A typical sample image is shown in Fig. 1(a). The electronic structure of Ta$_2$NiSe$_7$ is calculated using density functional theory (DFT) as implemented in the VASP package[26] and adopting the projector augmented wave potentials[27]. The Perdew-Burke-Ernzerhof functional within the gradient generalized approximation was used[28] and spin-orbit interaction was included. The Fermi surface was determined from the calculated bands using the software XCrySDen[29].

The Q1D structure of Ta$_2$NiSe$_7$ is illustrated in Fig. 1(b). Similar to FeNb$_3$Se$_{10}$,[30] the unit cell of Ta$_2$NiSe$_7$ consists of double rows of tantalum atoms (Ta1) in bicapped trigonal prismatic selenium coordination and the other double rows of tantalum atoms (Ta2) in octahedral selenium coordination; nickel atoms are in highly distorted octahedral coordination. As shown in Fig. 1(c, d), Ta$_2$NiSe$_7$ shows a metallic behavior with the current along the chain direction ($b$ axis) in the temperature range of 2~300 K for all magnetic fields between 0 and 14 T. A clear kink at around 61.8 K corresponds to the CDW transition temperature ($T_{CDW}$). The $T_{CDW}$ of our samples are among the highest in the literature.[23,31-33] MR for magnetic field applied along $b$ (MR$_{H\|I}$) and $c$ directions (MR$_{H⊥I}$) are both positive, as shown in the insets of Fig. 1 (c, d). Clearly, MR$_{H\|I}$ is small, while MR$_{H⊥I}$ is much larger and reaches up to 30% at low temperatures. In both cases, MR grows rapidly below the CDW transition. The high anisotropy in MR and its magnetic field orientation dependence indicate the dominant role of orbital MR, which has been discussed in detail in our earlier work.[24] The CDW transition is more evident in the first derivative of resistivity as shown in Fig. 1(e, f). It is important to note that upon increasing magnetic field with directions both perpendicular and parallel to the current, $T_{CDW}$ remains almost unchanged (insets in Fig. 1(e, f)). With the magnetic field of 14 T, an estimate of the energy scale associate with the magnetic field is $2μ_BH$, which compares to a significant fraction of about 1/3 of the CDW energy scale of $k_BT_{CDW}$. This field independence provides insight on this Q1D CDW system. An instructive CDW phase diagram in a magnetic field was established in an earlier calculation using random phase approximation of a Hubbard model.[20] The magnetic field dependence of $T_{CDW}$ showed a variety of behaviors: by increasing the imperfect nesting parameter, the CDW system is driven to pass from a regime with $T_{CDW}$ decreasing with the field to a regime with a nonmonotonical magnetic dependence of $T_{CDW}$. Among these, $T_{CDW}$ can indeed be field independent if the CDW nesting is imperfect,[20] which is a likely condition in $Ta_2NiSe_7$.

The main result is the angle-dependent MR shown in Fig. 2. Sample resistance is monitored while the external magnetic field is rotated in a crystallographic plane. Three different configurations were used, with $H$ rotating in a plane perpendicular to $c$ axis (Fig. 2(a, b)), parallel to $bc$ plane (Fig. 2(c, d), and parallel to $ac$ plane (Fig. 2 (e, f)), respectively, where angle $\theta$ is the inclination of $H$ from the principal axis in the plane. A dominant two-fold symmetry is seen in all configurations for $T$ above and below $T_{CDW}$. For $H \perp c$ and $H \parallel bc$, it is seen that the symmetry axis in $R(\theta)$ is the principal lattice axis, regardless of whether the system is in the CDW state or not. The resistivity in a rotating magnetic field reflects the symmetry of the underlying electronic states, which is determined by the lattice symmetry. $R$ shows maxima when $H$ is perpendicular to the current ($I \parallel b$) and minima when $H$ is parallel to the current, indicating that orbital MR dominates. It can be noted that the anisotropy shown in the polar plot of $\rho(\theta)$ is rather small for $T > T_{CDW}$, therefore a small misalignment of the magnetic field out of the rotating plane will lead to a small but noticeable background offset, with a magnitude of the order of 0.1% of the resistivity, as can be seen in Fig. 2(a, c, e).

The major finding of this work is the angle-dependent MR shown in Fig. 2 (e, f), where $H$ is in the $ac$ plane and is always perpendicular to the current. For $T > T_{CDW}$, $R(\theta)$ shows a similar behavior with that of the other two configurations: a two-fold symmetry with a principal lattice axis ($a$ axis in this case) being the symmetric axis in $R(\theta)$. Surprisingly, as the temperature goes below $T_{CDW}$, the symmetry axis shows a significant shift away from the position at high temperatures. We define a quantity $\theta_s$ which measures the deviation of the symmetric axis from that at high temperatures and track its temperature dependence, as shown in Fig. 3. Apparently, $\theta_s$ is essentially zero for $T > T_{CDW}$, and progressively increases upon cooling for $T < T_{CDW}$. The fact that such $\theta_s(T)$ behaivor only shows up for $H \parallel ac$ plane but not for the other two configurations implies its connection to the Q1D nature of the system.

The temperature dependence of $\theta_s$ is unexpected. First of all, since $Ta_2NiSe_7$ is diamagnetic,[23] there is no contribution from possible anisotropic magnetism. The system does not show significant lattice change in this temperature range either.[23] The symmetry change in $R(\theta)$ should reflect the corresponding change of the CDW states. Interestingly, we find that the shape of $\theta_s(T)$ resembles that of the temperature dependence of an order parameter. As shown in Fig. 3, a tentative fitting using the temperature dependence of a BCS gap function describes $\theta_s(T)$ reasonably well, which suggests that $\theta_s$ is associated with the order parameter of CDW, though the reason is not yet understood. On the other hand, it is useful to note that the CDW in Ta$_2$NiSe$_7$ involves two CDW wave vector of $2k_F$ and $4k_F$, each corresponding to the transverse displacement of Ni and Se2[32] and the longitudinal modulation of Ta2[33], respectively. The Major $2k_F$ CDW is found to exist only below 70 K, while the minor $4k_F$ CDW was found to persist up to 200 K.[23,33,34] The Former matches with the temperature dependence of the angle-dependent MR shown in Fig. 3, which indicates that the $2k_F$ CDW is responsible for the angle-dependent MR. This is consistent with the fact that $2k_F$ CDW is the major one, and there is no transport anomaly seen at above 200 K associated with the $4k_F$ CDW in our measurement and in earlier studies.[23]

Consider Ta$_2$NiSe$_7$ as a Q1D system, a qualitative analysis in Ref. [20] can be readily applied here. The magnetic field impacts the system through two mechanisms, the orbital effect and the Pauli effect. The former is characterized by the inverse magnetic length $q_0 \sim H\cos\theta$ (here, $\theta$ is defined as the inclination of $H$ from the transverse $c$ direction in the $bc$ plane). When $H$ is always rotating in $ac$ plane perpendicular to the current, the orbital effect is purely quantified by the term $H\cos\theta$. Therefore, it is anticipated that impact of orbital effect on the symmetry of $R(\theta)$ is determined by the relative angle $\theta$. This indeed is what we found for $R(\theta)$ in most cases but not for $H \parallel ac$ plane at $T < T_{CDW}$. The Pauli effect, on the other hand, sets in by the wave number $q_P = \mu_B H/v_F$, where the Fermi velocity $v_F$ can be anisotropic as a function of vector $k$. The behavior of $\theta_s(T)$ is very likely associated with the Pauli effect on the Fermi surface with anisotropic $v_F(k)$, particularly on the Fermi surface cross section perpendicular to $H$.

For a quantitative sense of the Fermi surface and the corresponding distribution of Fermi velocity $v_F$ of Ta$_2$NiSe$_7$, we have performed DFT calculation and obtained the Fermi surface in Fig. 4(a). Overall, the Fermi surfaces occupy a small fraction of the first Brillouin zone, consistent with its semimetallic nature. Two very narrow and flat Fermi surfaces dominate, both involves strong hybridization from the Se $p$ and Ta $d$ orbits. The flatness of Fermi surfaces is expected for the Q1D lattice. Part of the hole-like band labeled as FS1 and most of the electron-like one labeled as FS2 in Fig. 4(b, c) are more flat, which are compatible with the 1D nature observed in the angle-dependent MR. In particular, the flat section on FS1 is very likely nested by a nesting vector $q \simeq (0, 0.13b^*, 0)$, which is consistent with the previous finding that the major CDW occurred on the hole-like band. The nesting vector $q$ value is also close to an earlier value of $(0, 0.1b^*, 0)$ that reported earlier using a tight-binding calculation.[35] However, both $q$ values are significantly smaller than the value of $(0, 0.483b^*, 0)$ found by the X-ray and electron diffraction experiments.[23,33] The dominating Fermi surfaces show significant anisotropy, and thus provide an anisotropic $v_F(k)$, as illustrated quantitatively in Fig. 4(b, c). For a given $H$ direction, the anisotropic $v_F(k)$ determines the response due to the Pauli effect for that particular field direction and will change for different field directions because the cross section perpendicular to $H$ changes. This readily explains the anisotropy in $R(\theta)$ which conforms to the lattice symmetry, as those in Fig. 2 (a-e). However, the peculiar behavior of $\theta_s(T)$ in Fig. 3 with a rotating symmetry axis that does not find a corresponding lattice symmetry suggests additional mechanism. We propose that a shift of CDW nesting vector in the CDW-magnetic field phase diagram[20] is responsible. From a structural point of view, the inter-chain distance is around 3.0 Å between nearest Ni-Ta chains and 4.2 Å between nearest Ta-Ta chains, both are not large and can accommodate significant inter-chain coupling, as is supported by the charge density distribution from our DFT calculation. The isosurface plot in Fig. 4(d) shows significant charge density overlap between the neighboring chains, which is further seen in a representative contour plot of the charge density distribution at a plane cut perpendicular to (0, 0.5$b$, 0) in Fig. 4(e), which shows continuous nonzero charge density between neighboring chains. These features provide proper condition for the shift of nesting vector to occur[20]. It may also be useful to compare to the spin counterpart: the spin-density wave. In organic conductors, for example (TMTSF)$_2$PF$_6$, a set of field-induced-SDWs were indeed found, which in turn lead to remarkable quantum Hall effect in bulk crystal[36,37]. Such field-induced-SDWs are attributed to the shift of SDW nesting vector in a magnetic field to keep the carrier concentration constant. A natural question to ask is whether similar quantum behavior could exist in Ta$_2$NiSe$_7$, possibly at higher magnetic fields or lower temperatures. This sets up a tempting direction for the future work on this system. On the other hand, the considerably smaller wave vector $q$ obtained in our DFT calculations than that found by diffraction experiments may be an indication of more complexity of the CDW state. The possible role of the 4$K_F$ CDW in the low-temperature regime calls for further investigation.

Normally, insightful information regarding a CDW state is obtained through sophisticated scattering techniques, such as electron and X-ray diffractions. Here the establishment of the relation between the angle-dependent MR and the CDW states in the quasi-1D system makes the angle-dependent MR a sensitive and feasible transport probe of the quasi-1D CDW states. The angle-dependent MR is also naturally compatible with the investigation of possible field-induced phases and quantum CDWs, which can be useful for related material research.

To summarize, we found that the angle-dependent MR in the Q1D semimetal Ta$_2$NiSe$_7$ captures the Q1D nature of the system. It showed a surprising symmetry change in the CDW state, with a temperature dependence implying an intimate relation to the CDW order. We argue that this behavior is dominated by the Pauli effect, with the system featuring an anisotropic Fermi velocity distribution, and a shift of CDW nesting vector in a magnetic field. The former was corroborated by our first-principles calculation showing dominant flat Fermi surfaces with significant anisotropy in Fermi velocity, while the latter remains an open question for further test. Our findings show that the angle-dependent MR is a sensitive transport probe of the CDW, and may find more application in the general study of low-dimensional systems.

The authors are indebted to Anthony J. Leggett for inspiring discussions and suggestions. The authors have benefited from discussion with Antonio Miguel Garcia-Garcia and Mingliang Tian. The work performed at SJTU was supported by MOST (Grant No. 2015CB921104, 2016YFA0300500), NSFC (Grants No. 11804220, No. 91421304, No. 11574197 and No. 11474198) and the Fundamental Research Funds for the Central Universities, at Pennsylvania State University by the NSF (Grant No. EFMA1433378), at Tulane supported by the U.S. Department of Energy under Grant No. DE-SC0014208.

**Figure 1:**

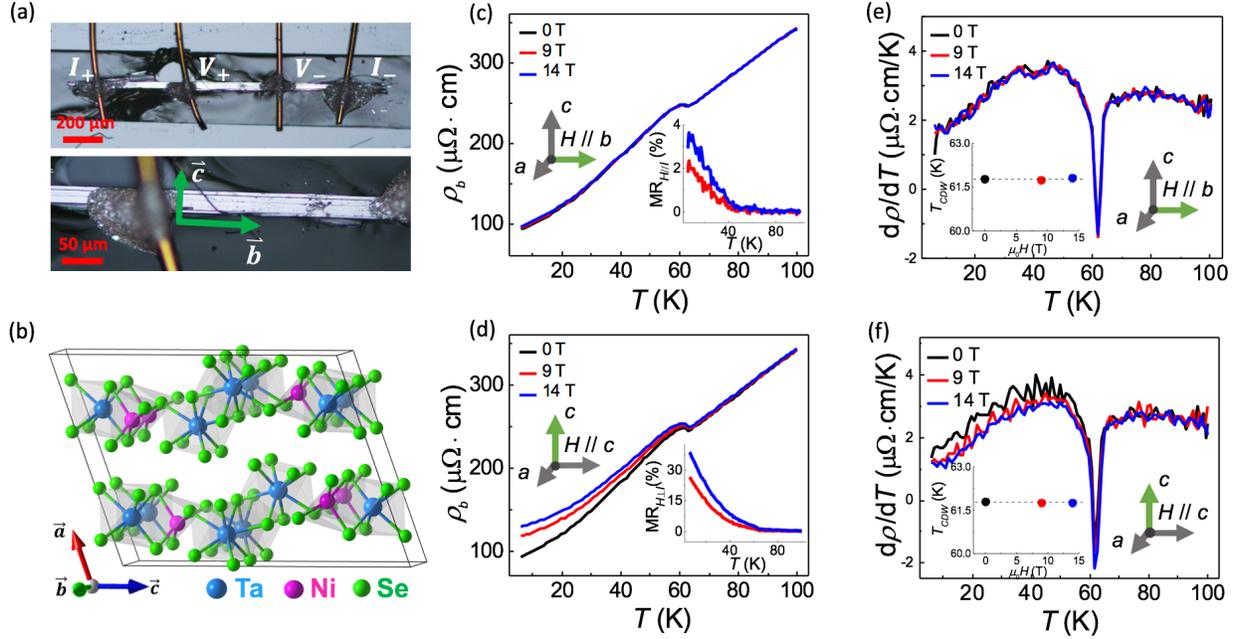

Fig. 1: (a) A typical optical image of Ta$_2$NiSe$_7$ single crystal used in the angle-dependent magnetoresistance measurements. The long edge of the sample is along the *b* axis. (b) A schematic of the lattice structure of Ta$_2$NiSe$_7$. (c, d) Temperature dependence of resistivity along *b* axis for *H // b* and *H // c* at various magnetic fields. Insets are magnetoresistance defined by $MR = [\rho_b(H) - \rho_b(0)]/\rho_b(0)$. The corresponding first derivative of resistivity $d\rho/dT$ are shown in (e, f). The sharp peak defines the CDW transition temperature, with its magnetic field dependence shown in the insets.

**Figure 2:**

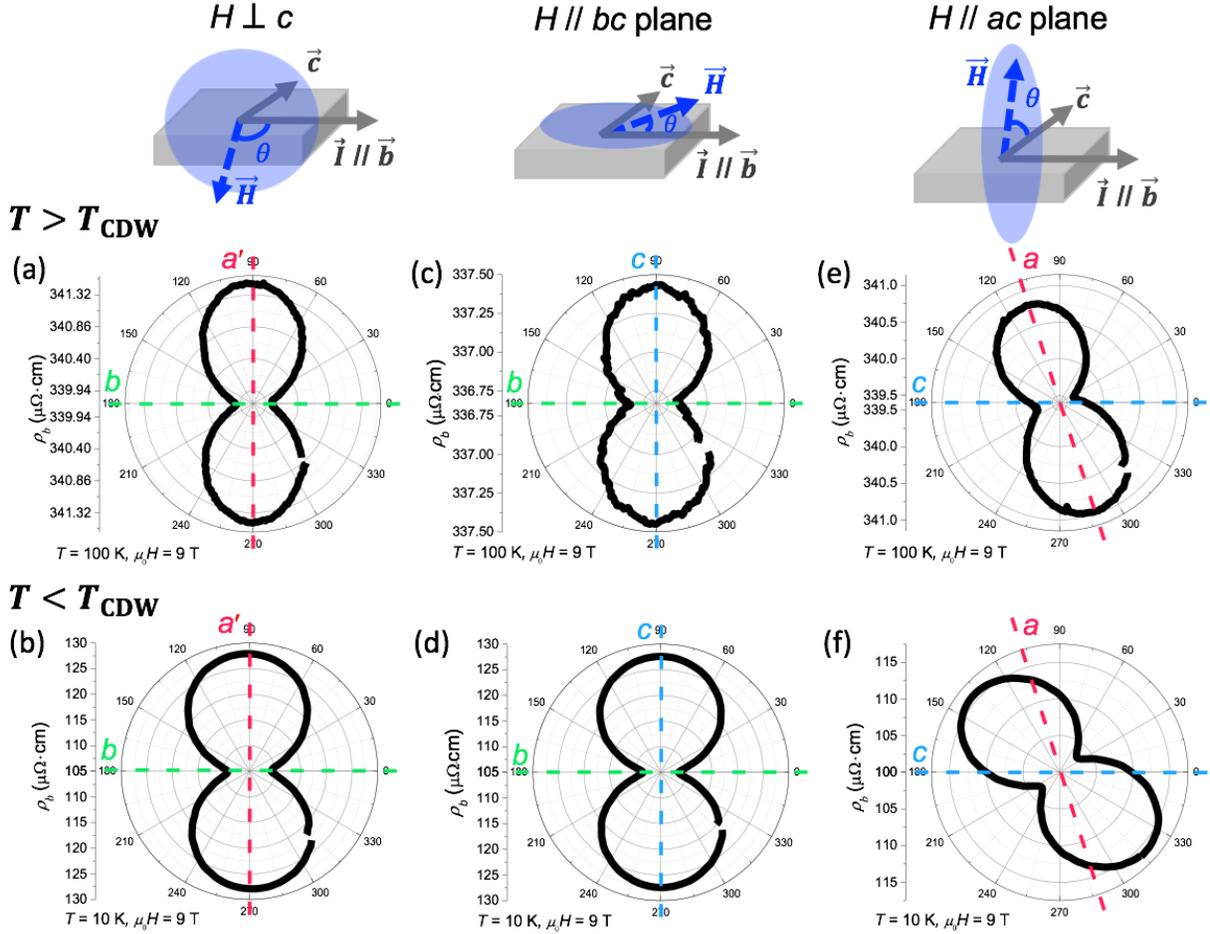

Fig. 2: The angle-dependent magnetoresistance of Ta$_2$NiSe$_7$ above and below $T_{CDW}$. Three different configurations are measured, with magnetic field rotating in a plane perpendicular to the *c* axis (a, b), parallel to the *bc* plane (c, d) and parallel to the *ac* plane (e, f), respectively. Dashed line in the polar plots represents the direction of the crystallographic axes *a*, *b* and *c*. Note that $a'$ represents the projection of the *a* axis to the plane perpendicular to *c* axis (as the angle between *a* and *c* axis is 108.8º).

**Figure 3:**

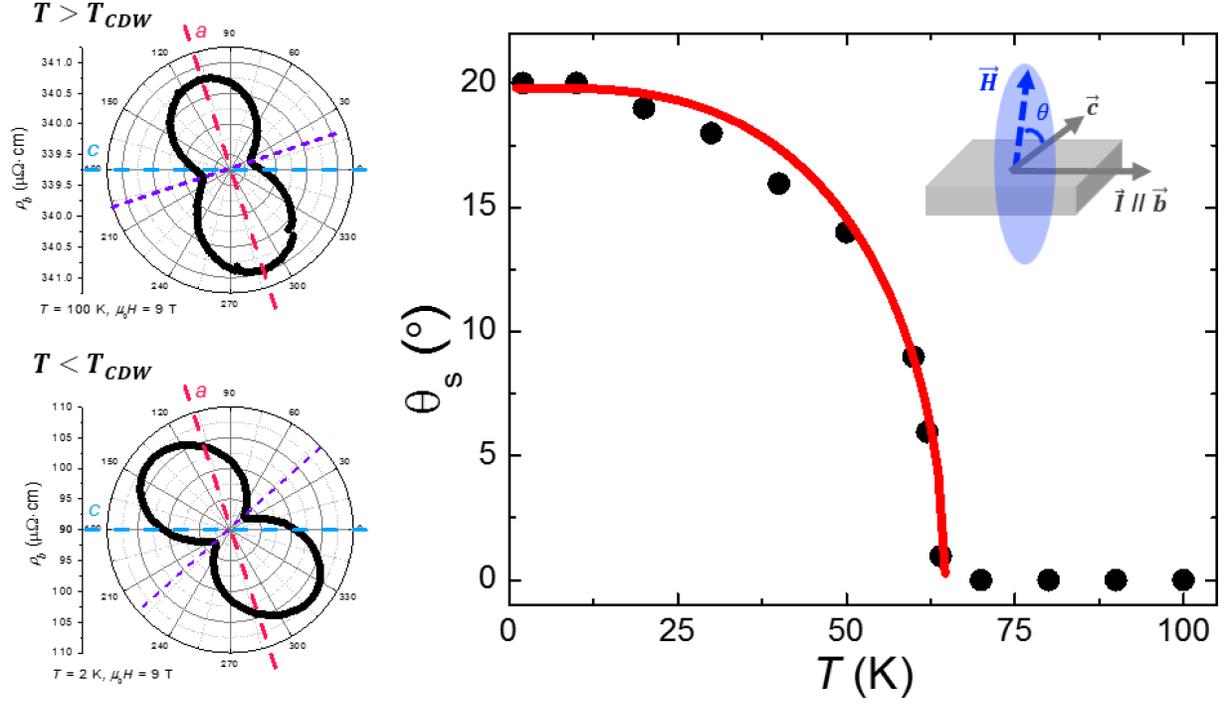

Fig. 3: For magnetic field rotating in the *ac* plane, the symmetry axis is denoted by the purple dash line. The deviation of the symmetric axis from that at high temperatures is defined as $\theta_s$. The temperature dependence of $\theta_s$ can be fitted using a BCS gap function (solid red line).

**Figure 4:**

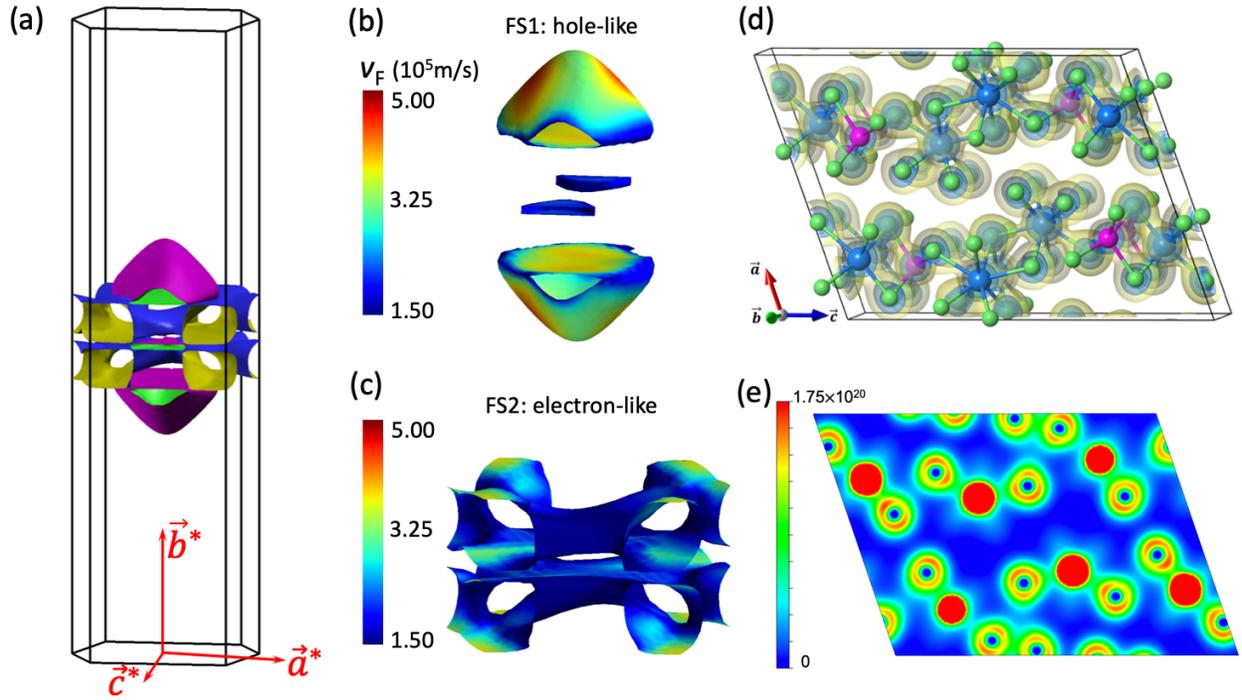

Fig. 4: Fermi surfaces of $Ta_2NiSe_7$ calculated using density functional theory. (a) The two primary Fermi surfaces plotted in the first Brillouin zone. $\vec{a}^*, \vec{b}^*$ and $\vec{c}^*$ are the reciprocal vectors. The two Fermi sheets are shown separately in (b) and (c) for clarity: a hole-like Fermi surface labeled as FS1, and an electron-like one labeled as FS2. The color scale in (b, c) represents the magnitude of Fermi velocity. (d) The isosurface plots of the charge density of $Ta_2NiSe_7$. The light yellow and light blue isosurfaces are for the charge densities of $5.64\times10^{19}$ cm$^{-3}$ and $1.35\times10^{20}$ cm$^{-3}$, respectively. (e) A representative contour plot of the charge density distribution at a plane cut perpendicular to (0, 0.5$b$, 0).